\documentclass[review,number,sort&compress]{elsarticle}
\usepackage{lineno}
\usepackage{amssymb}
\usepackage{subcaption}
\usepackage{graphicx}
\journal{Nuclear Instruments and Methods in Physics Research}

\begin{document}

\begin{frontmatter}



\title{Vibrations on pulse tube based Dry Dilution Refrigerators for low noise measurements}


\author[label1]{E. Olivieri}
%\ead{emiliano.olivieri@csnsm.in2p3.fr}
\address[label1]{CSNSM, Univ. Paris-Sud, CNRS/IN2P3, Universit\'e Paris-Saclay, 91405 Orsay, France}

\author[label2]{J. Billard}
\author[label2]{M. De Jesus}

\author[label2]{A. Juillard}

\address[label2]{Univ Lyon, Universit\'e Lyon 1, CNRS/IN2P3, IPN-Lyon, F-69622, Villeurbanne, France}

\author[label3]{A. Leder}
\address[label3]{Massachussets Institute of Tehnology, Laboratory for Nuclear Science, 77 Massachusetts Avenue
Cambridge, MA 02139-4307}

%\author[6]{---add co-authors---}

\begin{abstract}
Dry Dilution Refrigerators (DDR) based on pulse tube cryo-coolers have started to replace Wet Dilution Refrigerators (WDR) due to the ease and low cost of operation. However these advantages come at the cost of increased vibrations, induced by the pulse tube.
%Companies provide decoupling solutions to mitigate vibrations, which are more or less effective, showing vibration performance obtained under undocumented conditions, poorly described and detailed, and often contradictory.
In this work, we present the vibration measurements performed on three different commercial DDRs. 
We describe in detail the vibration measurement system we assembled, based on commercial accelerometers, conditioner and DAQ, and examined the effects of the various damping solutions utilized on three different DDRs, both in the low and high frequency regions. Finally, we ran low temperature, pseudo-massive (30 and 250 g) germanium bolometers in the best vibration-performing system under study and report on the results.
\end{abstract}

\end{frontmatter}


\section*{Introduction}
\label{}
Due to helium shortage and increasing price of liquid helium, in the last decade, research groups performing experimental physics  at low temperatures  have begun to replace the usual Wet Dilution Refrigerators (WDR) with pulse tube-based Dry Dilution Refrigerators (DDR). The success of DDRs relies on the low-cost and ease of operation. In particular, the high level of automation of the gas handling systems and the lack of liquid helium bath allow for a nearly autonomous cool down and running. However, pulse tubes induce vibrations which are so far the most serious drawback  of this technology \cite{gm_vibrations,vib_free_4K_stage}.
Indeed, vibrations can drastically affect the results of experiments as in the case of Scanning Tunnelling Microscopy, Johnson Noise Measurements and Bolometers \cite{article_placement_rm,article_placement_rm2}.

The ultimate goal for DDR technology is to provide, through an efficient vibration decoupling system, a low temperature and  low vibration environment as good as the one obtained with WDRs. Throughout this paper we assume that, in first approximation, running a DDR fridge with its pulse tube turned OFF is equivalent in terms of vibrations to running a WDR.\\

In this work, we propose a vibration measurement standard (\S \ref{sec:chain}), built with market-based components, that allows for a rigorous and unambiguous comparison between vibration levels of DDRs, at room temperature. We set three vibration limits to classify systems as  {\it noisy}, {\it typical} and {\it quiet}. We report on vibration measurements on three (four) different DDRs (setups) and draw conclusions on their vibration performances (\S \ref{sec:acc}).

Finally(\S \ref{sec:bolo}), we show how vibration levels as measured with accelerometers compare with bolometers, highlighting the need for vibration levels below $\mathrm{10\ \mu g}$ to operate these correctly.\\

\section{Description of the DDR units under study}
\label{sec:DDR}
\begin{figure}[h!]
\begin{center}
\includegraphics[width=0.8\linewidth,keepaspectratio]{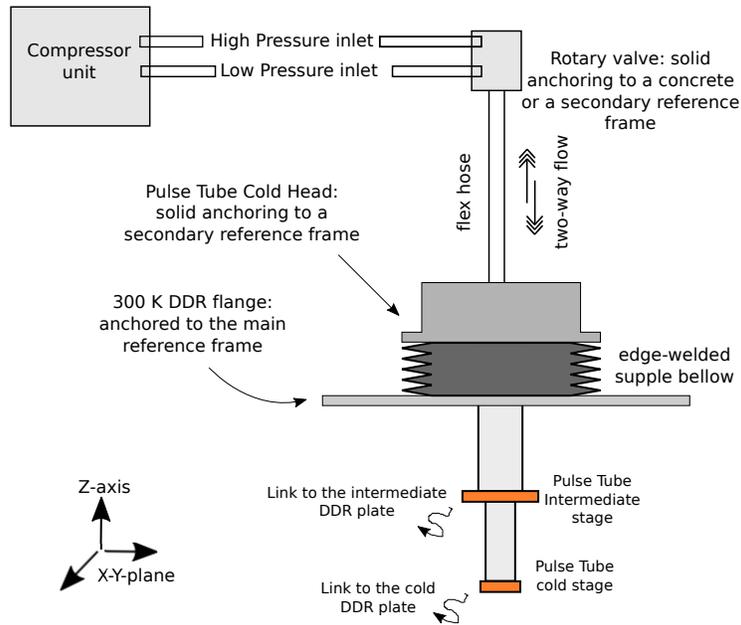}
\caption{General scheme depicting how to efficiently filter the vibrations injected by the pulse tube on the DDR unit. Orientation and positioning of the rotary valve and flex hose do matter.}
\label{schematic}
\end{center}
\end{figure}

Here below we list the three (four) DDRs (setups) under study and describe the various vibration damping solutions utilized by each one (Fig.~\ref{schematic}).
\begin{itemize}
\item[-]{{\bf Hexadry Standard (Hex std):} produced by Cryoconcept, it is the standard model of the Hexadry Hexagas $\texttrademark$  series \cite{cryoconcept_website}. It is equipped with a PT410 Cryomech pulse tube with a remote rotary valve. The pulse tube cold head is tightly fixed onto the 300 K flange, without any dedicated vibration decoupling system. The pulse tube intermediate and cold stages are thermally coupled to the cryostat intermediate (50 K) and cold (4 K) plates via low pressure gas-exchangers (Hexagas $\texttrademark$) to avoid any mechanical contact and hence reduce the propagation of vibrations down to the various cold stages of the fridge. No special care was devoted to the positioning of the remote motor, which was held on the main DDR unit frame \cite{cryoconcept_website}. The unit was installed at the {\it Institut de Physique Nucl\'eaire de Lyon} (IPNL) and devoted to detector R\&D for the \emph{EDELWEISS} dark matter search experiment \cite{edelweiss}}.
\item[-]{{\bf Hexadry Ultra Quiet Technology (Hex UQT):} it is exactly the same aforementioned DDR unit but upgraded with the UQT (Ultra Quiet Technology $\texttrademark$) option. This option is especially conceived to provide  a  low vibration environment at low temperatures. It consists in a mechanical decoupling of the pulse tube head from the rest of the cryostat via an edge-welded supple bellow \footnote{The edge-welded bellow employed has an elastic constant of 30 N/mm along the $z$ axis whereas the radial constant is of 2200~N/mm.}. A few mm-thick neoprene O-ring is installed between the bellow and the 300 K flange to cut out high frequency vibrations.  A solid secondary frame, physically separated from the main one, is firmly mounted on the ceiling and rigidly holds the pulse tube head \cite{cryoconcept_website}. The rotary valve may be mounted on the ceiling to further decouple from the cryostat.
An analogous system, {\bf Hex UQT (STERN)}, has kindly been set at our disposal by Cryoconcept and Bar-Ilan University (Israel) \cite{STERN} to study the reproducibility of the vibration performances with respect to the unit and installation site. For this unit, the pulse tube head and rotary valve were both mounted on a secondary frame, separated from the cryostat main frame.}
\item[-]{{\bf Oxford Triton 400 (Triton):} produced by Oxford Instruments \cite{oxford_website}, the system is especially conceived to provide a low temperature, low vibration experimental environment. This design utilizes an edge-welded bellow to insulate the vibrations coming from the pulse tube head and provides thermal contacts between the pulse tube stages and the cryostat intermediate (50 K) and cold (4 K) plates via supple copper braids. The system comes mounted on a single solid frame (main frame). All the different dilution unit cold plates down to the coldest (10 mK plate) are rigidly triangulated. The unit uses the same pulse tube as the Cryoconcept models with a remote rotary valve option. For our experimental studies we evaluated a system installed at the Laboratory for Nuclear Science at MIT, currently used for ongoing \emph{CUORE/CUPID} detector R\&D \cite{cuore_cupid} .}\end{itemize}
\label{fig:setup}

In this work we will see that the vibrations induced by the pulse tube can be transmitted to the dilution unit both via the cold head (300 K pulse tube flange) and the cold stages. Hence, an efficient vibration damping solution must take both into account.\\

The gravitational wave experiment \emph{CLIO} \cite{clio_experiment_pt}  first realized a 4 K non vibrating cold plate cryostat, by decoupling the pulse tube cold head with an edge-welded supple bellow and utilizing supple copper braid thermal links between the pulse tube stages and the intermediate (50 K) and cold (4 K) cryostat plates. Since, this decoupling solution is commonly adopted in dry refrigerators.

Nevertheless, the  \emph{CLIO} experiment observed residual vibrations on the cryostat plates; it demonstrated these were transmitted mainly by the mechanical thermal links and negligibly from the edge-welded bellow. This prompted Cryoconcept to opt for thermal couplings through gas-exchangers \footnote{A gas-exchanger consist of two annular, entangled counter-radiators. The fixed radiator is accommodated on the cryostat intermediate (cold) plate whereas the counter-radiator is tightly fixed on the pulse tube stage(s). This latter sits inside the fixed radiator with a gap of few mm, without any mechanical link. Low pressure helium gas establishes the thermal link between the two counter-radiators. This gas-exchanger technique is a trademark of Cryoconcept.} through the Hex UQT $\texttrademark$ technology.

A special care must be applied in choosing and dimensioning the edge-welded bellows to decouple the pulse tube cold head; in fact, bellows efficiently damp vibrations along their axial direction $z$, whereas they perform poorly along the radial direction $r$ \footnote{The stiffness coefficient $k_z$ of the edge-welded bellow along $z$ direction is much smaller than the radial one $k_r$.}. Fortunately, though pulse tube vibrations are not negligible along the radial direction, the majority of these are along the axial direction \cite{measure_xyz}.

\section{Description of the measurement system}
\label{sec:chain}

To measure the vibrations at the Mixing Chamber (10~mK cold plate) of the different DDRs and setups, we selected and set up a measurement system standard, composed of a high sensitivity {\it PCB-393B04} seismic accelerometer (PCB Piezo-electronics, typical sensitivity of 1~V/g in the 1~Hz-750~Hz frequency region), a {\it PCB-480E09} signal conditioner and a 16-bit National Instrument DAQ-6218. The measurement chain has been carefully chosen to evidence the residual low level of vibrations injected by the pulse tube down to $\mathrm{0.2\ \mu g\sqrt{Hz}}$, in the  $\mathrm{1\ Hz-1\ kHz}$ frequency range. 

Two other accelerometers were tested: {\it PCB-351B41}(cryogenic) and {\it Kistler-8762A} (3-Axes). They have been rejected because their intrinsic noise was too large to appreciate vibrations at the required level. 

We mounted the accelerometer on the Mixing Chamber ($10\ mK$ plate), allowing it to sense along the vertical and radial directions. For reading the signal, we used an anti-tribo-electric coaxial cable, tightly fixed to the rigid structures of the DDRs (to avoid spurious signal induced by the stress or vibrations of the cable). A leak-tight electrical feedthrough was used to connect this latter cable to the conditioner which sat outside the cryostat. We performed the measurements with the OVC (Outer Vacuum Chamber) under vacuum to prevent the accelerometer from picking up the acoustic environmental noise through air.\\

All measurements have been performed at room temperature, for three reasons: 1) the lack of any low budget easy-to-handle cryogenic accelerometer with sufficiently low intrinsic noise; 2) to first order, we assume that the room temperature acceleration measurements are representative of the vibration level at low temperatures. Indeed, no large difference between the 300 K and 4 K values of the elastic constant $k$ and Young's modulus $E$ is observed for stainless steel and copper \cite{Baron_book}, which are the main materials used for the rigid structures of the DDR units; 3) room temperature measurements can be performed rapidly by any user, with much less constraints as those at low temperatures.
\section{Acceleration and displacement: results and discussion}
\label{sec:acc}
\subsection{Accelerations}
\label{sec:accelarations}
We measured the acceleration of the Mixing Chamber (10 mK cold plate) of the three (four) DDR units (setups) via the acquisition chain described in the previous section. The signals from the conditioner were sampled at 16 bits, 10~kHz, over a $\pm1\ V$ range.

We performed a Fast Fourier Transform (FFT) analysis  using Hanning windowing over 5~s time windows. We trace the acceleration power spectral density $PSD_a$, as a function of the frequency, for each time window. The spectra were then averaged according to:
\begin{equation}
PSD_a=\sqrt{1/N\sum_{i_{th}=1}^{N}(PSD_a)_{i_{th}}^2}\qquad \mathrm{ [g/\sqrt{Hz}]}
\end{equation}
where $N$ is the total number of time windows (about 25 for all measurements).

For convenience, we first define three relevant vibration levels in the $PSD_a$ domain: a) {\it typical} ($\mathrm {1\times10^{-5}\ g/\sqrt{Hz}}$) for low noise measurements. Several low temperature, low noise experiments will be able to run without any issue within this level. We have a global convergence of the bolometer community toward this value. b) {\it noisy} ($\mathrm {1\times 10^{-4}\ g/\sqrt{Hz}}$). This is the upper "acceptable" limit of vibration for low and ultra low temperatures ($\mathrm {T<10\ mK}$). At this level, vibrations can impact the base temperature reached by the DDR Mixing Chamber. c) {\it quiet } ($\mathrm {1\times 10^{-6}\ g/\sqrt{Hz}}$) which represents a difficult level to achieve, as it requires special installations, as anechoic chambers and laminar air-flow isolators.\\
\begin{figure}[h!]
\begin{center}
\begin{subfigure}[b]{1\textwidth}
\includegraphics[width=1\linewidth,keepaspectratio]{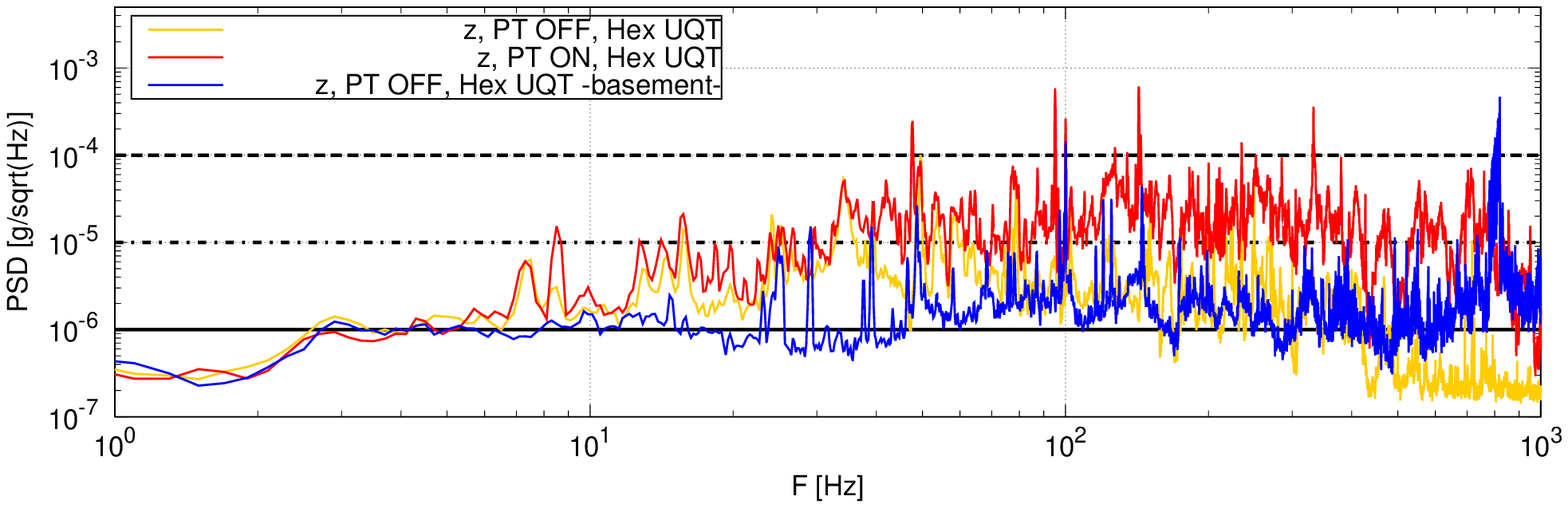}
\end{subfigure}
\begin{subfigure}[b]{1\textwidth}
\includegraphics[width=1\linewidth,keepaspectratio]{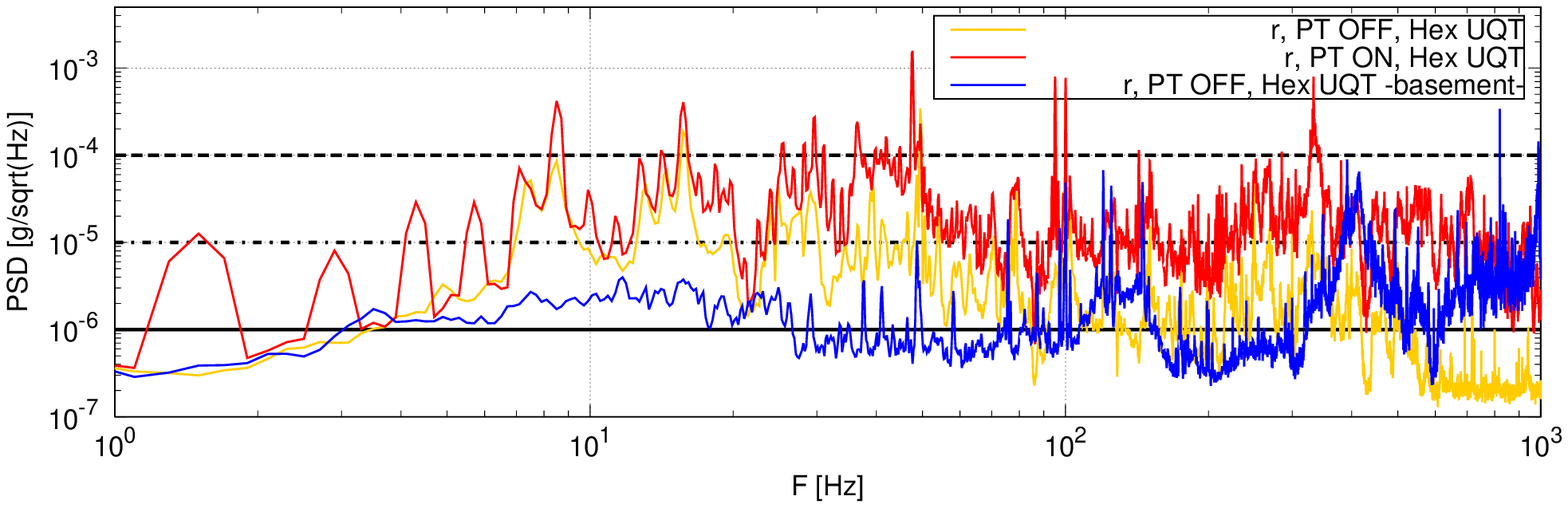}
\end{subfigure}
\end{center}
\caption{Hex UQT $PSD_a$ as a function of the frequency, along the $z$ (top) and $r$ (bottom) directions, for pulse tube ON and OFF. Three notable vibration levels are also traced: a) {\it typical} vibration level, for low noise measurements (black dot-dashed-line) b) {\it noisy} vibration level (black dashed-line) and c) {\it quiet} vibration level (black solid-line). The floor $PSD_a$ has been also carefully acquired and traced (-basement-)}
\label{acceleration_psd}
\end{figure}
To facilitate any further discussion, we also define two relevant frequency regions as follows:
REG1) {\it mechanical frequency range}, from 1~Hz up to 40~Hz. It represents the region where the pulse tube mechanically induces displacements of the cryostat (vibrating at the pulse tube fundamental frequency and first harmonics). These movements stem mostly from the elongation of the flex hose connecting the pulse tube cold head to the rotary valve and from the displacement of the pulse tube cold stages. 

Indeed, the pulse tube experiences pressure variations between 9 and 18 bars every cycle and the flex hose behaves as a piston. A possible solution to reduce the contribution due to the movements of the flex hose is to replace it with a rigid pipe. Large benefits in terms of vibrations from this configuration have been observed, with the remote motor tightly held on a concrete block/wall~\cite{rigid_pipe_private_communication}.
We also noted that the frame holding the cryostat can present resonant frequencies in this range, hence a special care should be devoted to its design.
REG2) {\it acoustic frequency range}, from 40~Hz up to 1~kHz, which is  the frequency region where ``acoustically audible noise'' populates the vibration measurements. Gas flowing through the pulse tube corrugated pipes and flex hose typically contributes in this range as it generates a whistle-like audible noise. Moreover, in this frequency region, the OVC acts as a resonating bell which then injects these acoustic vibrations into each DDR cold plates through the rigid structures of the cryostat.\\

Fig. \ref{acceleration_psd} reports the acceleration measurements for the Hex UQT, for the pulse tube turned ON/OFF, along the axial $z$ (top) and radial $r$ (bottom) directions, respectively. For this specific setup we observe almost no difference along the $z$ direction in REG1. However, due to the transversal stiffness of the edge-welded bellow we see vibrations at the fundamental pulse tube frequency (1.4~Hz) and harmonics, along the radial directions (bottom). We could mitigate the transversal vibrations on this specific setup by mounting the rotary valve as recommended in Fig.~\ref{fig:setup}, with the flex hose aligned along the z-axis.
In the acoustic region REG2, we clearly observe for both the $z$ and $r$ directions the pulse tube noise.\\
\begin{figure}[h!]
\begin{center}
\includegraphics[width=1\linewidth,keepaspectratio]{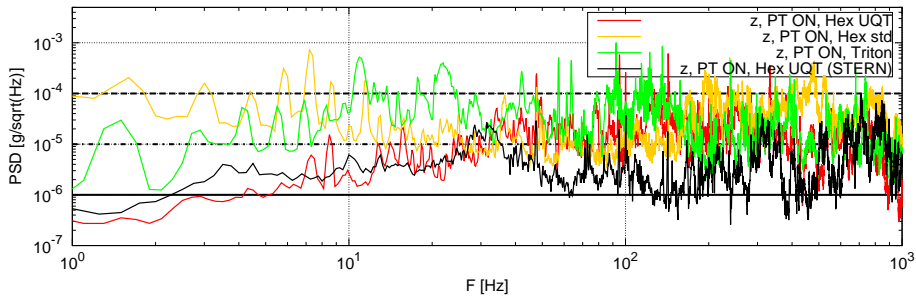}
\end{center}
\caption{$PSD_a$ for different DDR units (pulse tube on), as a function of the frequency, along $z$ direction.}
\label{comp_acceleration_psd}
\end{figure}
Fig.~\ref{comp_acceleration_psd} compares the vibration spectra along the axial direction, for the three (four) units (setups) under study. The Hex UQT showcases the best vibration damping, capable of reducing the pulse tube-induced vibrations up to two orders of magnitude. We point out the reproducibility of the vibration performances of the Hexadry UQT technology, highlighted by the black and red solid lines. Both the Triton and the Hex std show in REG1 pulse tube fundamental and harmonic peaks, although the Triton is more favorable. The measurements on the Hex std show that efficient vibration damping can only be achieved by combining both the gas exchanger technology and the thorough decoupling of the pulse tube cold head with respect to the 300~K flange.

In REG2 vibrations are strongly related to the acoustic environmental noise. For the Hexadry UQT (STERN), a special care was devoted to acoustically isolate the OVC and reduce the ``audible acoustic noise'' contribution. For this reason, it largely outperformed the other units. In particular, it showed a vibration level as good as for pulse tube turned OFF.

\subsection{Displacements}
As experiment performances may be more easily interpreted in terms of displacements, for the sake of completeness, we now discuss our results in terms of displacements. The displacement power spectral density $PSD_d$ can be derived from the acceleration $PSD_a$ by double integration  in the frequency domain, as follows:
\begin{equation}
PSD_d(f_i)=\mathrm{\frac{(9.81\ m/s^2)}{(2\cdot \pi\cdot f_i)^2}\cdot PSD_a(f_i) \qquad \left[m/\sqrt{Hz}\right]}
\label{eq:PSD}
\end{equation}
where $f_i$ corresponds to the frequency bins. 

\begin{figure}[h!]
\begin{center}
\begin{subfigure}[b]{1\textwidth}
\includegraphics[width=1\linewidth,keepaspectratio]{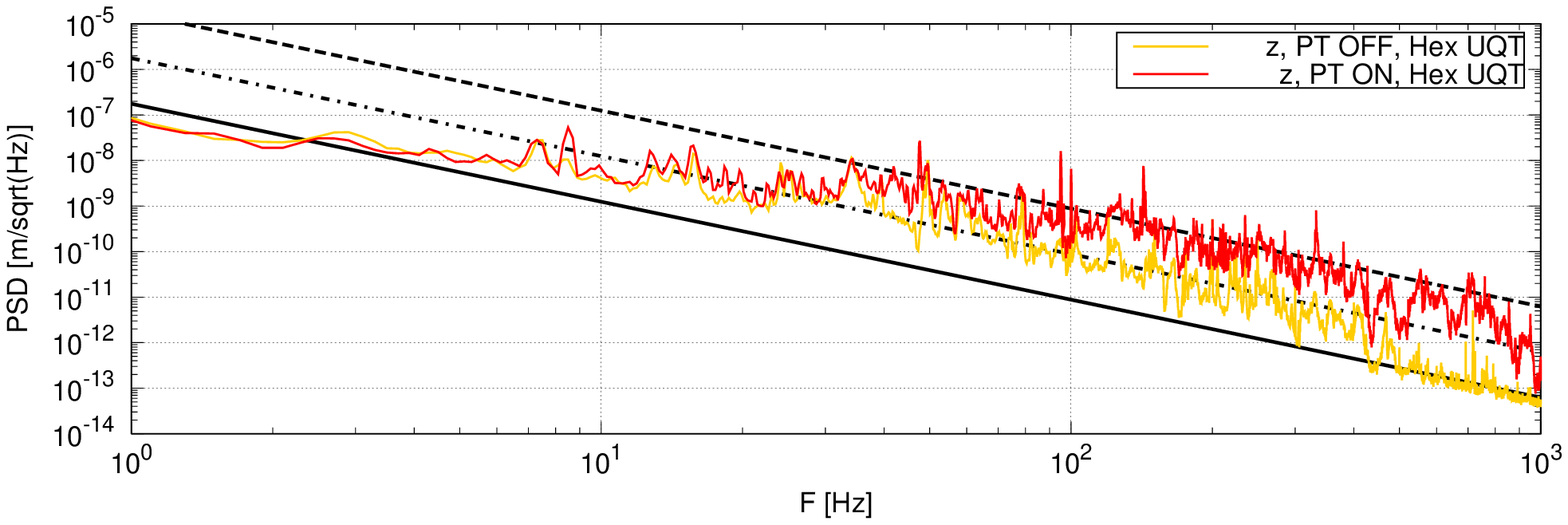}
\end{subfigure}
\begin{subfigure}[b]{1\textwidth}
\includegraphics[width=1\linewidth,keepaspectratio]{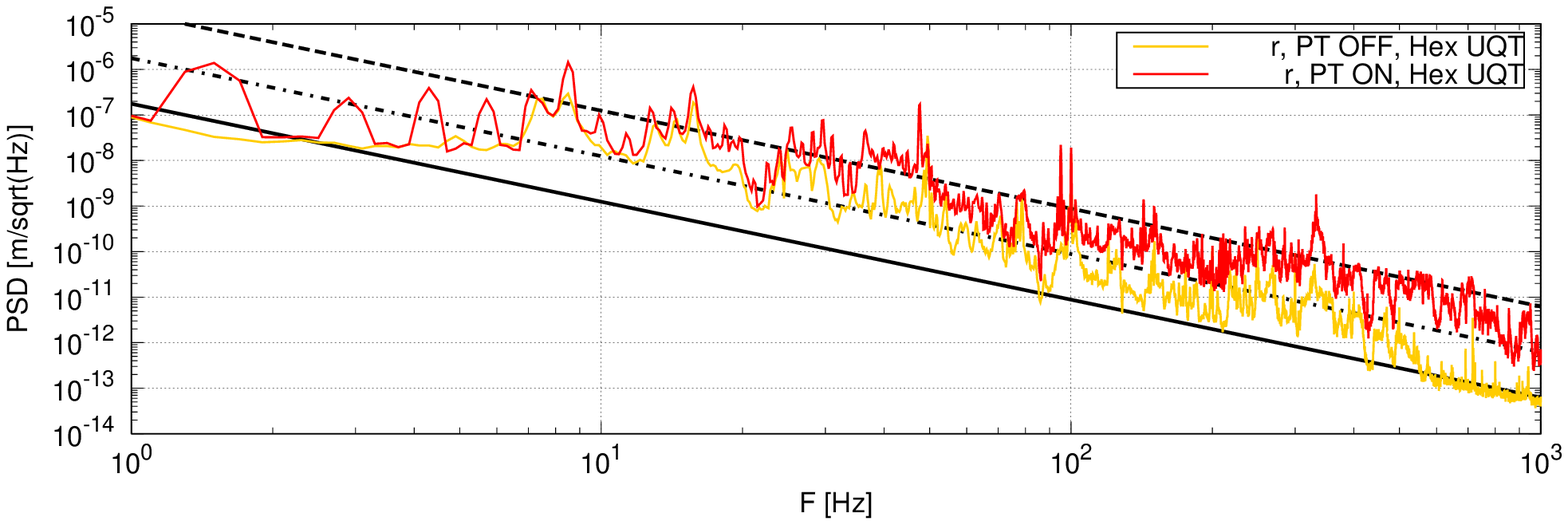}
\end{subfigure}
\end{center}
\caption{Hexadry UQT displacement power spectral densities as a function of the frequency, for pulse tube turned ON and OFF, along the axial $z$ and radial $r$ directions. The three notable vibration levels have been propagated and traced in the displacement plot.}
\label{displacement_psd}
\end{figure}

\begin{figure}[h!]
\begin{center}
\includegraphics[width=1\linewidth,keepaspectratio]{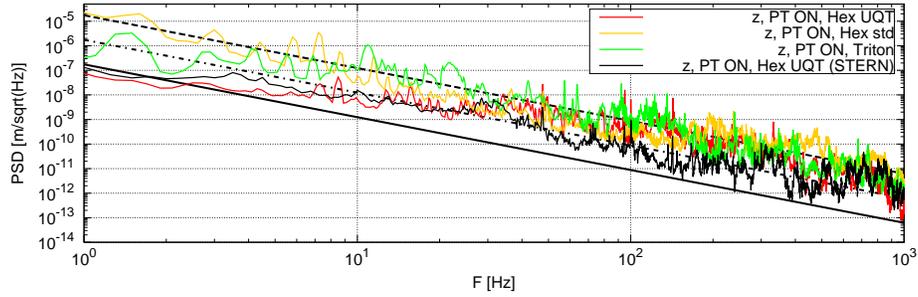}
\caption{Hexadry UQT displacement power spectral densities as a function of the frequency, for the three (four) different setups.}
\label{compare_displacement_psd}
\end{center}
\end{figure}

Fig.~\ref{displacement_psd} shows the $PSD_d$ for the Hex UQT, along the axial $z$ and radial $r$ directions, for the pulse tube turned ON and OFF, whereas Fig.~\ref{compare_displacement_psd}  compares the displacements ($PSD_d$) for the three (four) units (setups). Due to the $1/f^2$ dependence, the low frequency modes can easily dominate the displacement measurements. 
To better compare the displacement levels of the setups, we calculate for each of them the $RMS$ displacement over the REG1 and REG2 frequency regions, according to following formula (derived from Parseval's theorem):
\begin{equation}
RMS\big|_{f_l}^{f_h}=\sqrt{\sum_{f=f_l}^{f_h}(PSD_d)^2\Delta f}\qquad \mathrm{[m]}
\end{equation}
where $\Delta f$ is the discrete frequency step and $f_l$, $f_h$ are the limit of the frequency range ($\Delta f=\frac{1}{tw}$,where $tw=5$ s is the time window chosen to perform the FFT analysis.).
\label{}
Tab. \ref{rms_table} reports the results and the intrinsic $RMS$ noise limit of our measurements.
Looking at the results in REG1, we see how the edge-welded decoupling system combined with the gas-exchanger technique (Hex UQT) is effective in reducing mechanical vibrations. In contrast, using only the gas-exchanger technique (Hex std) is inefficient.

The UQT technique is quite reproducible, as shown by the comparison between the two UQT setups. Furthermore, the main/secondary frames solution adopted in the Hex UQT (STERN) performs well, though anchoring mechanically the pulse tube head and rotary valve to the ceiling or a concrete block yields better results.

The vibration reduction system adopted on the Triton is less effective than the Hex UQT system. However, it is difficult to conclude if the residual displacements for this unit stem from the cold stage braid links or from the cold head.

We moreover see that the displacements along the radial direction $r$ are  one order of magnitude larger than those along $z$. Efforts are needed to mitigate the transmission of the vibrations along the radial direction and reach the limit already achieved with the Hex UQT along the $z$ direction.

The displacement results show that no correlation exists between the $RMS$ displacements of the two frequency regions. The better performing unit in REG2, the Hex UQT (STERN), has a displacement level which compares with the Hex std, which is by far the worst performing in REG1.
\begin{table}[htp]
\begin{center}
\resizebox{\textwidth}{!}{%
\begin{tabular}{|l|l|c|c|}
\hline
&\bf DDR unit&\bf REG1: $\mathrm {1\ Hz<f<40\ Hz}$&\bf REG2: $\mathrm{40\ Hz<f<1\ kHz}$\\
\hline
1.&Hex std, $\rightarrow z$&$14.4\ \mu $m ($6.7\ \mu $m)&$13\ $nm ($7\ $nm)\\
2.&Triton, $\rightarrow z$&$2.37\ \mu $m ($0.33\ \mu $m)&$56\ $nm ($19\ $nm)\\
3.&Hex UQT (STERN), $\rightarrow z$&$0.111\ \mu $m ($0.110\ \mu $m)&$3\ $nm ($3\ $nm)\\
4.&Hex UQT, $\rightarrow z$&$0.071\  \mu $m ($0.066\ \mu $m)&$25\ $nm ($22\ $nm)\\
5.&Hex UQT, $\rightarrow r$&$1.22\ \mu $m ($0.300\ \mu $m)&$116\ $nm ($22\ $nm)\\
\hline
6.&Int. noise limit (digit)&$52\ $nm&$1\ \AA$\\
\hline
\end{tabular}%
}
\end{center}
\caption{Comparison of the RMS noise of the three (four) units (setups), evaluated in REG1 and REG2, for pulse tube ON and OFF. Note how the acoustic insulator dramatically reduces the $RMS$ noise in the REG2, by comparing two Hex UQT setups.}
\label{rms_table}
\end{table}
\section{Pseudo-massive, high impedance NTD sensor bolometers}
\label{sec:bolo}
In this section we report how the performance and noise of two pseudo-massive (30~g and 250~g) germanium bolometers compare with the vibration levels of various DDR systems. Both detectors were equipped with high impedance NTD (Neutron Transmutation Doped) thermal sensors \cite{ntd_article}. The 30 g detector was first operated in the Cryoconcept Hex std cryostat at IPNL. After observing a strong correlation between the pulse tube vibrations on the detector's performances and operating temperatures, we upgraded the cryostat with the Ultra Quiet Technology vibration reduction system and transformed it into a Hex UQT. The 30~g bolometer was again tested, utilizing strictly the same read-out system and cabling. We subsequently ran a high sensitive, 250~g germanium bolometer.

\subsection{Bolometer and setup description}
The two bolometers were rigidly anchored to the 10~mK cold  plate and electrically connected to a room temperature  read-out electronics via anti-tribo-electric constantan-copper coaxial cables. Special care was devoted to the thermalisation of the cables at each DDR cold stages. We measured the temperature of the 10 mK cold  plate via a  calibrated $RuO_2$ resistive thermometer \cite{full_range} and regulated via an electrical P.I.D.-controlled heater.
%The cryostat is equipped with a 1st and 2nd cold stage copper thermal shields at 50 K and 3.6 K respectively, and a Still thermal shield, which seats around 850 mK.There is no 50 mK or 10 mK thermal shield.

We utilized a CUORE-like read-out system \cite{cuoricino_electronics}, which consists in purely DC,  low noise, high stability amplifiers providing an overall gain of 2400 (tunable), combined with a 4 poles, 2 kHz low-pass Bessel filter. The analog output of the electronics was sampled at 16 bits, 10 kHz, over a $\pm$10 V dynamic range (NI-6218 DAQ). Overall, the read-out shows an intrinsic voltage noise of 4~nV/$\mathrm {\sqrt{Hz}}$, above 1 Hz and up to the Bessel cutoff frequency. 

\subsection{Resistance vs. Temperature curves}
The detection principle of the considered bolometers using a $NTD$ thermometer is based on the fact that a particle interaction with the germanium absorber increases the absorber temperature of a few micro-kelvins and induces a variation of the resistance of the thermal sensor. As the latter is current biased, we then observe a voltage signal across the sensor. The resistance of such sensors as a function of temperature follows Mott-Anderson law \cite{mott-anderson}:
\begin{equation}
\label{mott_law}
R(T) = R_0\exp\left({\sqrt{T_0/T}}\right)
\end{equation}
where $R_0$ depends mainly on geometrical factors and $T_0$ is related to the germanium doping level. 
\begin{figure}[h!]
\begin{center}
\includegraphics[width=1\linewidth,keepaspectratio]{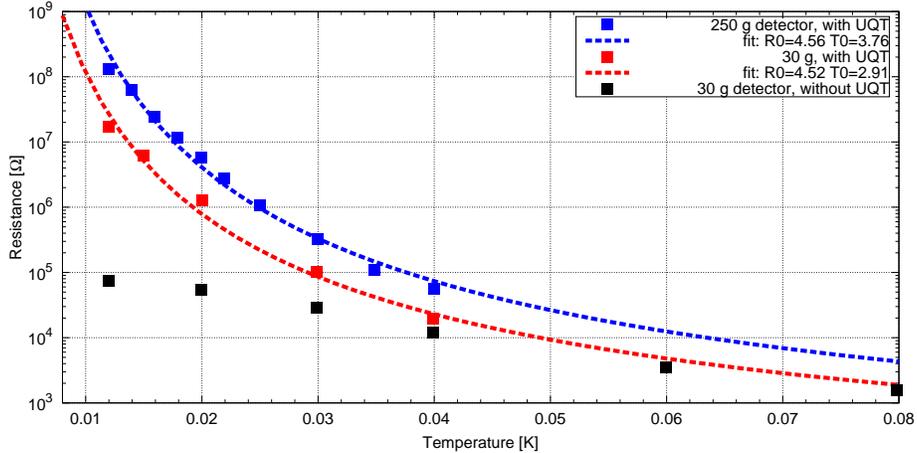}
\caption{Resistance of the $NTD$ germanium thermal sensors of both bolometers, as a function of the 10 mK cold plate temperature, for the 30 g detector before (black symbols) and after (red symbols) the upgrade from Hex std to Hex UQT. The characteristic curve of the 250 g detector (blue symbols) is also traced, for comparison. Dashed lines show the fits of the data with Eq.~(\ref{mott_law}).}
\label{RTcurve}
\end{center}
\end{figure}
Fig. \ref{RTcurve} shows the characteristic curves (resistance vs. temperature) of the 30~g bolometer obtained before/after the upgrade (black/red symbols) of the cryostat. Before the upgrade, the bolometer temperature levels off around 100~k$\Omega$, which corresponds to a temperature of about 30~mK (extrapolated using the measurements after the upgrade), whereas the Mixing Chamber temperature approached the 12~mK.
The down-conversion of mechanical vibrations into heat via several mechanism, {\it e.g.} friction between the bolometer absorber and the clamps holding it, results in a constant power injection and hence, in the heating of the bolometer. 

Thanks to the UQT upgrade, both bolometers recovered the expected characteristic curves, falling into agreement with Eq. \ref{mott_law} as shown by the fit. 
\subsection{Bolometer noise spectra}
In this section we study the impact of pulse tube-induced vibrations on the noise of a bolometer. We focus on the 250~g germanium detector because of its increased sensitivity compared to the 30~g detector, thanks to its improved thermal sensing design~\cite{ThermalLTD}.  We operated the detector at a fixed temperature of 18~mK which corresponds to the standard operating temperatures of the \emph{EDELWEISS} experiment ~\cite{edelweiss}.

To characterize the role of the vibrations toward the bolometer thermal noise (signal power spectral density) and disentangle the contribution of the cabling microphonics \cite{cabling_vibrations}, we operate the bolometer in two configurations, together with pulse tube ON/OFF:
\begin{itemize}
\item[a)] No polarisation current: in this configuration the bolometer thermal sensitivity is null and we can solely test piezo-electric and tribo-electric contributions (microphonics) to the bolometer noise due to cabling. Tribo/piezo-electricity produce charge (current) noise, which translates into voltage noise through the NTD impedance, which was of 12~M$\Omega$.
\item[b)] Optimally polarised: at about 1~nA polarisation current, the NTD impedance lowers to $\sim 8$~M$\Omega$ and the bolometer is maximally sensitive to thermal variations and energy deposit. With a sensitivity of 200 nV/keV it allows us to probe the effect of the pulse tube vibrations and their down-conversion to heat into the absorber.
\end{itemize}
\begin{figure}[h!]
\begin{center}
\includegraphics[width=1\linewidth,keepaspectratio]{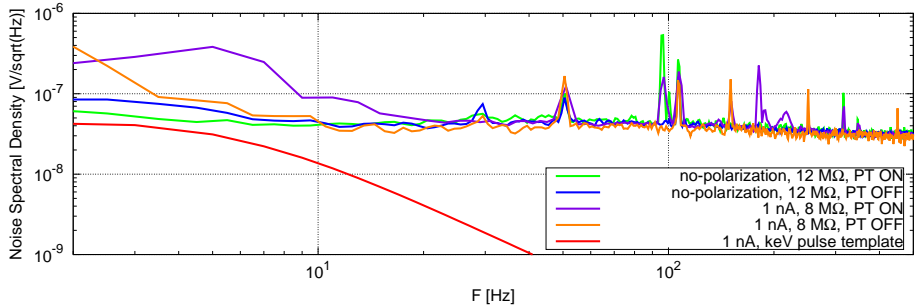}
\caption{Noise Power Spectral Densities ($PSD_V$), for pulse tube ON and OFF under no bias (blue and green) and under the optimal 1 nA bias current (orange and purple). Also shown is the sensor response to a 1 keV event in the frequency domain (red). These results were obtained using the 250 g detector that has the highest sensitivity of 200 nV/keV and at 18 mK.}
\label{dionysos_psd}
\end{center}
\end{figure}
\label{}
The resulting noise power spectral densities as a function of the frequency are reported in Fig.~\ref{dionysos_psd}. The green (PT-ON) and blue (PT-OFF) curves correspond to configuration a), whereas the purple (PT-ON) and orange (PT-OFF) curves correspond to configuration b). In red we also show  the signal response of the detector, normalized to a 1 keV energy event. All the noise power spectra have been computed using a 500 ms time window, to avoid pile-up events, and traced up to  500~Hz. For the case where we operated the detector in mode b), due to a particle event rate of about 2~Hz and an intrinsic bolometer signal decay-time of about 60~ms, an additional chi-square cut was applied to select pure noise samples, as any decaying tail from an event can mimic a 1/f-like noise and therefore bias our noise power spectral density.\\

A small 50~Hz noise (european AC power supply frequency) and higher order harmonics pollute the noise spectra. The slight contribution at 30 Hz comes from a pick-up of the data acquisition system.

By comparing pulse tube ON/OFF measurements in configuration a), we observe no difference on the overall noise spectra: vibration-induced microphonics of the cabling has a negligible contribution to the bolometer noise.

However, by comparing pulse tube ON/OFF in configuration b), even though no significant additional noise contributions are seen in the 30~Hz-500~Hz range, we do observe an excess of noise at low frequencies. From optimal filter theory \cite{enss}, we evaluate the energy resolutions to be of 2.5~keV and 1.7~keV (RMS), for the pulse tube ON/OFF, respectively. This difference in noise is due to the fact that the bolometer studied is particularly sensitive to low frequencies (below 20 Hz), with the dominant noise contributions stemming from pulse tube residual vibrations, most likely along the radial $r$ directions. These results have triggered investigations to additionally mitigate radial vibrational modes, as discussed in \S \ref{sec:accelarations}.

\section{Conclusions and Recommendations}
Pulse tube-induced vibrations have a dramatic effect on the operation of massive and pseudo-massive bolometers at cryogenic temperatures.

We showed how we have designed and set up a vibration measurement system based on commercial accelerometers, conditioner and DAQ, well suited to measure the accelerations in low noise environment.
We have studied in details the vibrations level on the Mixing Chamber (10~mK plate) of three (four) different DDR units (setups), with large differences observed in terms of vibrations and displacements. The most effective vibration mitigation technology combines the  decoupling of the pulse tube head via edge-welded bellow together with gas-exchangers, as implemented on the Cryoconcept Hex UQT model. We confirmed the importance of a secondary frame, separated from the main DDR frame, to tightly hold the pulse tube head and the rotary valve.\\

Of all the technologies we examined, the Cryoconcept Hex UQT can bring the vibration level down most effectively in both the low and high frequency regions and allows to run massive and pseudo-massive bolometers.

Improvements are possible on DDRs to further reduce vibrations at the 10~mK cold plate by installing an additional secondary 10~mK floating plate, suspended via spring-loaded, mass-damped wires and thermally linked with supple, high conductivity copper braids \cite{floating_plate}.

\section{Acknowledgements}
The results of this work were only made possible through the collaborative effort of several partners. We wish to especially thank Cryoconcept, which granted us access to several DDR units at their factory before delivery and provided us with valuable assistance in upgrading our setups.
We also address special thanks to the cryogenic group of the {\it SPEC-IRAMIS-CEA} laboratory, led by P. Pari, and the associated mechanical workshop for the technical discussion and valuable mechanical realizations. Finally, we thank P. Camus and M. Pyle for their fruitful discussions about new vibration reduction strategies.


\end{document}